\documentclass[runningheads,a4paper]{llncs}

\usepackage{amssymb}
\usepackage{graphicx}
\usepackage{comment}
\usepackage{framed}
\usepackage{amsmath}
\usepackage{color}
\usepackage[ruled,vlined]{algorithm2e}

\newcommand{\ra}{\rightarrow}

\newcommand{\la}{\leftarrow}

\newcommand{\set}[1]{\{#1\}}                      

\newcommand{\tup}[1]{\langle #1\rangle}            

\def\qed{\hfill{\qedboxempty}      
  \ifdim\lastskip<\medskipamount \removelastskip\penalty55\medskip\fi}

\def\qedboxempty{\vbox{\hrule\hbox{\vrule\kern3pt
                 \vbox{\kern3pt\kern3pt}\kern3pt\vrule}\hrule}}

\def\qedfull{\hfill{\qedboxfull}   
  \ifdim\lastskip<\medskipamount \removelastskip\penalty55\medskip\fi}

\def\qedboxfull{\vrule height 4pt width 4pt depth 0pt}

\newenvironment{noteac}{\begin{leftbar} \textbf{Note
      (Andrea):}}{\end{leftbar}}

\renewcommand{\emptyset}{\varnothing}

\newcommand{\isa}[1]{\mathit{ISA}}

\renewcommand{\paragraph}[1]{\textbf{#1}}

\newcommand{\red}[1]{\textcolor{red}{#1}}

\newcommand{\mc}[1]{\mathcal{ #1}}

\newcommand{\nit}[1]{{\it #1}}

\newcommand{\boxtheorem}{\hfill $\blacksquare$\vspace*{0.2cm}}
\newcommand{\wboxtheorem}{\hfill $\Box$\vspace*{0.2cm}}

\newcommand{\ignore}[1]{}

{\vskip \abovedisplayskip \refstepcounter{lemmaA-counter}%
\noindent {\bf Lemma A.\arabic{lemmaA-counter}.}}%

\newcounter{lemmaA-counter}
\newcommand{\datalogpm}{Datalog$^\pm$}

\author{Mostafa Milani$^1$ \and Andrea Cal\`\i$^{2,3}$ \and Leopoldo
  Bertossi$^1$} \institute{
  \begin{minipage}[t]{13em}
    \centering
    $^1$School of Computer Science\\
    Carleton University, Canada
  \end{minipage}
\hspace*{2em}
  \begin{minipage}[t]{15em}
    \centering
    $^2$Dept of Computer Science\\
    Birkbeck, Univ.~of London, UK
  \end{minipage}\\[1em]
  \begin{minipage}[t]{24em}
    \centering
    $^3$Oxford-Man Institute of Quantitative Finance\\
    University of Oxford, UK
  \end{minipage}\\[1em]
  \texttt{\{mmilani}, \texttt{bertossi\}{@}scs.carleton.ca}\\
  \texttt{andrea{@}dcs.bbk.ac.uk} }

\begin{document}
\sloppy
\pagestyle{plain}

\mainmatter  

\title{A Hybrid Approach to Query Answering \\under Expressive Datalog$^\pm$}

\maketitle

\begin{abstract}

  \datalogpm{} is a family of ontology languages that combine good
  computational properties with high expressive power.  \datalogpm{} languages
  are provably able to capture many relevant Semantic Web
  languages. In this paper we consider the class of weakly-sticky (WS)
  \datalogpm{} programs, which allow for certain useful forms of joins in rule bodies as well as extending the well-known class of
  weakly-acyclic TGDs.  So far, only nondeterministic algorithms were known
  for answering queries on WS \datalogpm{} programs.  We present novel
  deterministic query answering algorithms under WS \datalogpm{}.  In
  particular, we propose: \textsl{(1)} a bottom-up grounding algorithm based on a query-driven chase, and
  \textsl{(2)} a hybrid approach based on transforming a WS program into a
  so-called sticky one, for which query rewriting techniques are known.  We
  discuss how our algorithms can be optimized and effectively applied
  for query answering in real-world scenarios.


\end{abstract}


\section{Introduction}
\label{sec:introduction}

The \datalogpm{} family of ontology languages~\cite{cali12-jws}, which extends
Datalog with explicit existential quantification, has been gaining importance
in the area of ontology-based data access (OBDA) due to its capability of
capturing several conceptual data models and Semantic Web languages as well as
offering efficient query answering services in many variants relevant for
applications.

The core feature of \datalogpm{} languages are the so-called existential rules
(a.k.a.~\emph{tuple-generating dependencies} or~\emph{TGDs}).  Such rules
allow the inference (entailment) of new atoms from an initial set of ground
atoms, typically a database, through the \emph{chase}
procedure~\cite{deutsch,fagin,johnson}.
For example, consider the rule $\forall X\forall Y\;r(X,Y) \rightarrow \exists
Z\, s(X,Z)$ (in the following we shall omit universal quantifiers, keeping
only the existential ones) and a database $D$ constituted by a single atom
$r(a,b)$.  A chase step will generate (notice that $X$ and $Y$ correspond to
$a$ and $b$ respectively in this step) the atom $s(a,\zeta_1)$, where $\zeta$
is a \emph{labelled null}, that is, a placeholder for an unknown value; notice
that the constant $b$ is lost in this step as it doesn't appear in the new
atom.  In general the chase may not terminate.
Answering a conjunctive query (CQ) $q$ under a database $D$ and a set of TGDs
$\Sigma$ amounts to computing all atoms entailed by $D \cup \Sigma \cup q$
(where $q$ is seen as a TGD) or, equivalently, evaluating $q$ over all atoms
entailed by $D \cup \Sigma$.

Consider the following simple example.  Let $\Sigma =
\set{\sigma_1,\sigma_2}$; $\sigma_1$ is the rule $\mathit{emp}(X) \ra \exists
Y\,\mathit{rep}(X,Y)$, asserting that each employee reports to someone;
$\sigma_2$ is the rule $\mathit{rep}(X,Y) \ra \mathit{mgr}(Y)$, asserting that
anyone to whom someone else reports is a manager.  Now assume $D =
\set{\mathit{mgr}(\mathit{ann}), \mathit{emp}(\mathit{joe})}$.  Let us
consider the conjunctive query $q_1$ defined as $q_1(W_1) \la
\emph{rep}(W_1,W_2)$, where $q_1$ is now a new predicate (we use the head
predicate~$q_1$ for convenience, but any predicate name would suit).  This
query, which is itself a TGD\footnote{Here we adopt the usual notation for
  conjuntive queries, where the head appears on the left-hand side.},
which we denote with~$\delta$, asks for all those who report to someone.
Clearly the plain evaluation $q_1(D)$ over~$D$ returns no answer ($q_1(D) =
\emptyset$), as $D$ says nothing about who reports to whom.  However, we are
to reason on atoms entailed by $D \cup \Sigma \cup \delta$; therefore, we
apply the chase of $D$ w.r.t.~$\Sigma\cup \delta$.  By applying~$\sigma_1$ to
$\mathit{emp}(\mathit{joe})$ we obtain $\mathit{rep}(\mathit{joe},\zeta_1)$,
where $\zeta_1$ is a labeled null (in logical terms this means $\exists
Y\,\mathit{rep}(\mathit{joe},Y)$); then by applying~$\sigma_2$ to
$\mathit{rep}(\mathit{joe},\zeta_1)$ we obtain $\mathit{mgr}(\zeta_1)$; no
more atoms can be entailed by the chase.  Now, if we evaluate~$q_1$ on all
entailed atoms, thus computing all atoms entailed by~$D \cup \Sigma \cup
\delta$, we get the answer~$\tup{\mathit{joe}}$; this is because $D \cup
\Sigma \cup \delta$ entails that $\mathit{joe}$ reports to someone: $D \cup
\Sigma \cup \delta \models q_1(\mathit{joe})$.
Consider now the query $q_2$ defined as $q_2(W_1) \la \emph{mgr}(W_1)$, asking
for all those who are managers.  In this case, evaluating $q_2$ over all atoms
entailed by $D \cup \Sigma$ returns the answer $\tup{\mathit{ann}}$; notice
that $D \cup \Sigma$ does not entail the answer $\tup{\zeta_1}$ for~$q_2$;
this because~$\zeta_1$ is just a placeholder for an unknown value --- we know
that the manager to whom~$\mathit{joe}$ reports exists, but we do not know who
he (or she) is.

Conjunctive query answering under general TGDs is undecidable; languages of
the \datalogpm{} family impose therefore restrictions on the form of rules so
as to guarantee decidability and certain computational problems, most
prominently, conjunctive query answering.

\emph{Guarded} and \emph{weakly-guarded} \datalogpm{} (the latter generalising
the former) were the first decidable \datalogpm{} languages, inspired by
guarded logic and characterised by the presence of a \emph{guard atom} in each
rule that contains all variables of that rule~\cite{cali12}.
The \emph{sticky} \datalogpm~\cite{cali12} language was introduced to capture
a ``proper'' notion of \emph{join} in rules, that is, the occurrence of
variables in two distinct atoms of a rule body in the absence of a guard for
that rule.  \emph{Weakly-sticky (WS)} \datalogpm{} extends sticky
\datalogpm{} by also capturing the well-known
class of \emph{weakly-acyclic} rules~\cite{fagin}.
As an example, consider the following set of rules $\Sigma$, where some body
variables are \emph{marked} (by a hat sign, e.g.~$\hat{X}$; see~\cite{cali12};
notice that $X$ and $\hat{X}$ are the same variable) as the result of a
procedure that identifies occurrences of variables (the marked ones)
corresponding, in the chase procedure, to a value that can eventually be lost
in some subsequent chase step.
\[ \arraycolsep=15pt
\begin{array}{rl c rl}
  \sigma_1:\hspace{-8mm}&v(X)~ \rightarrow~\exists Y\;r(X,Y). & \hspace{-10mm} & \sigma_3:\hspace{-8mm}&r(X,\hat{Y}),r(\hat{Y},Z)~ \rightarrow~p(X,Z).\\
  \sigma_2:\hspace{-8mm}&p(\hat{X},\hat{Y})~ \rightarrow~\exists Z\;p(Y,Z).  & \hspace{-10mm} & \sigma_4:\hspace{-8mm}& p(\hat{X},Y),p(Y,Z)~ \rightarrow~t(Y,Z).
\end{array}
\]
A set of TGDs is \emph{sticky} if, for each rule, there is no marked variable
in that rule that appears more than once in the body of the same rule.
Intuitively, stickiness can be defined by means of the following (semantic)
property: during a chase step according to a rule $\sigma$, each value
corresponding to a variable appearing more than once in $\sigma$ is not lost
in the chase step, and it is also never lost in any subsequent step involving
atoms where it appears.  Notice that the set of rules $\Sigma$ above is not
sticky, as easily seen.

%


%

Weak acyclicity is defined using the notion of \emph{rank} of a position,
i.e. a predicate attribute. In a position of finite (resp., infinite) rank,
the number of labelled nulls that can appear in the chase procedure is finite
(resp., infinite). In the set of TGDs $\Sigma$ above, $\Pi_F(\Sigma)=\{v[1],
r[1], r[2]\}$ contains the positions with \emph{finite rank} and
$\Pi_\infty(\Sigma)=\{p[1], p[2],t[1],t[2]\}$ those with \emph{infinite rank}.
A \datalogpm{} program is \emph{weakly-acyclic} if all positions have finite
rank.  The set~$\Sigma$ of TGDs above is not weakly-acyclic.


A set of TGDs is WS if, for each TGD, every marked variable
that appears more than once in the body also appears at least once in a
finite-rank position.  This notion generalizes both stickiness and acyclicity,
because the stickiness condition applies to variables that appear \emph{only}
in positions with infinite rank.  Notice that $\Sigma$ above is WS.
Specifically, in $\sigma_3$, the repeated variable $Y$ appears in positions
$r[1]$ and $r[2]$, which are in $\Pi_F(\Sigma)$.  In $\sigma_4$, $Y$ is
repeated not marked.

To answer a conjunctive query $q$ under a set $\Sigma$ of TGDs (or other types
of ontological rules) and a database~$D$, two main approaches were proposed in
the literature: \emph{grounding} (or \emph{expansion}) and \emph{query
  rewriting}.  In the grounding approach, variables are suitably replaced by
constants (or nulls) in the body of a rule, so that the head of the rule
yields a (ground) atom that is entailed by the program.  The aforementioned
chase is in fact a grounding procedure.  The grounding allows the computation
of all atoms entailed by $D \cup \Sigma$, onto which $q$ can be then
evaluated.  In the rewriting approach, the query~$q$ is rewritten, according
to $\Sigma$, into another query~$q_R$ (possibly in another language different
from that of~$q$), so that the correct answers can be obtained by
evaluating~$q_R$ directly on $D$.

The rewriting approach is usually considered more efficient than the
grounding because in the former only the query is manipulated, according to
the rules, while the data is left unchanged; on the contrary, the grounding
approach requires the expansion of the given data, whose size is normally
much larger than that of the query and of the rules.

CQ answering can be done in polynomial time in data for WS programs. However,
so far, no non-trivial deterministic algorithm for CQ answering has been
devised for WS~\datalogpm.  In this paper we devise algorithms for
the efficient implementation of conjunctive query answering under
WS \datalogpm.  Our contributions are as follows.
\begin{enumerate} \itemsep-\parsep
\item We propose a bottom-up technique, based on grounding, which is a variant
  of the chase procedure, and relies on a terminating chase-like procedure
  that is \emph{resumed} a number of times that depends on the query to be
  evaluated.  Once the procedure is resumed a sufficient number of times for
  the query, the same query is evaluated together with the result of the
  procedure, that is a set of (ground) rules, yielding the correct answers to
  the query under the given WS set of TGDs.
\item
  We propose a \emph{hybrid} approach between grounding and rewriting as
  follows.  First, with a novel algorithm, we transform the given set~$\Sigma$
  of TGDs into another, all whose positions in $\Pi_F$ become of rank~$0$
  (that is, positions where only constants can appear in the chase or
  grounding).  Next, certain variables in such positions are \emph{grounded},
  that is, they are replaced by selected constants of the given instance.  The
  obtained program is a sticky program, for which a well-known query rewriting
  technique for CQ answering can be applied.  The rewriting (a union of CQs)
  is finally evaluated on the initial instance.
\end{enumerate}
Both techniques we propose yield algorithms for CQ answering that are of the
same data complexity as the lower-bound complexity of CQ answering under
weakly-stick programs.
Moreover, both our algorithms, unlike the one for WS \datalogpm{}
in~\cite{cali12}, are deterministic.  The advantage of the first,
pure-grounding technique is that we can pre-compute off-line a ground Datalog
program (possibly containing nulls, treated as constants), which serves for
answering every query up to a certain number of variables
by simply evaluating it on
the minimum model of the above Datalog program (in fact, a relational
instance).  The second, hybrid algorithm relies on a partial grounding, again
computed off-line, as well as on an on-the-fly rewriting of every query; the
advantages of this approach are that~\textsl{(a)} the grounding is
query-independent and generally much smaller than a complete grounding;
\textsl{(b)} the final step consists of a mere evaluation of a union of CQs on
the given database.  Notice that in both our approaches the last step can be
performed by executing an SQL query, then offering the possibility of taking
advantage of optimizations of RDBMSs.
Several optimization strategies are possible for these algorithms (this is
ongoing work).  We argue that our techniques set the basis for efficient CQ
answering under expressive \datalogpm{} languages such as
WS~\datalogpm.  A full version of this paper is available
at~\cite{MiCB16}.
%


\vspace{-2mm}
\section{Preliminaries}
\label{sec:preliminaries}
\vspace{-3mm}

In this section, we review some basic notions which we use in the paper.
%

\vspace{-2mm}

\subsection{Basic definitions}

\vspace{-2mm}

We assume an infinite universe of data constants $\Gamma_C$, an infinite set
of (labeled) nulls $\Gamma_N$, and an infinite set of variables $\Gamma_V$
(used in rules and queries). \ignore{Different constants represent different
  values (unique name assumption), while different nulls may represent the
  same value. We assume a lexicographic order on $\Gamma_C\cup\Gamma_N$, with
  every symbol in $\Gamma_N$ following all symbols in $\Gamma_C$. } We denote
by uppercase letters (e.g. $X,Y,Z$) variables, while $\vec{X}$ is a sequence
of variables $X_1,\ldots,X_k$ with $k \geq 0$. We use the same notation for
sets of variables.  A relational schema $\mc{R}$ is a finite set of relation
names (or predicates). A position $p[i]$ identifies the $i$-th argument of a
predicate $p$.


A homomorphism is a structure-preserving mapping $h:\Gamma_C\cup \Gamma_N \cup \Gamma_V \rightarrow
\Gamma_C \cup \Gamma_N \cup \Gamma_V$ such that $c\in \Gamma_C$ implies
$h(c)=c$.
For atoms and conjunctions of atoms, we denote by
$\Pi$-homomorphism a homomorphism that is the identity on the terms that
appear in a set $\Pi$ of positions.

A \emph{tuple-generating dependency} (or \emph{TGD}, also called
\emph{existential rule}) $\sigma$ on a schema $\mc{R}$ is a formula
$p_1(\vec{X}_1), \ldots, p_n(\vec{X}_n) \rightarrow \exists
\vec{Y}\;p(\vec{X},\vec{Y})$ in which $p,p_1,\ldots,p_n$ are predicates in
$\mc{R}$ and $\vec{X} \subseteq \bigcup \vec{X}_i$ and $\bigcup \vec{X}_i$ are universal variables that are implicitly quantified. We denote by ${\it
  head}(\sigma)$ and ${\it body}(\sigma)$ the head atom $p(\vec{X},\vec{Y})$
and the set of the body atoms $p_1(\vec{X}_1), \ldots, p_n(\vec{X}_n)$ of
$\sigma$, respectively.  Variables in~$\vec{Y}$ are called \emph{existential
  variables}.  A rule is \emph{ground} if its terms are in
$\Gamma_C\!\cup\!\Gamma_N$.

An instance $D$ for $\mc{R}$ is a (possibly infinite) set of atoms with
predicates in $\mc{R}$ and arguments from $\Gamma_C \cup\Gamma_N$. A database
$D$ is an instance that contains only atoms with arguments
from $\Gamma_C$. The \emph{active domain} of a database $D$ denoted
by ${\it active}(D)$ is the set of constants that appear in $D$.

A rule $\sigma$ is satisfied by an instance $I$, written $I \models \sigma$,
if the following holds: whenever there exists a homomorphism $h$ such that
$h(\nit{body}(\sigma)) \subseteq I$, then there exists a homomorphism $h'$ as
an extension of $h$ that maps existential variables of $\sigma$ into terms in $\Gamma_C\cup\Gamma_N$, such that $h'(\nit{head}(\sigma))\subseteq I$. An
instance $I$ satisfies a set $\Sigma$ of TGDs, denoted $I \models \Sigma$, if
$I \models \sigma$ for each $\sigma \in\Sigma$.

A conjunctive query (CQ) has the form $q(\vec{X}) \leftarrow
p_1(\vec{X}_1),...,p_n(\vec{X}_n)$ where $p_1,...,p_n$ are predicate names in
$\mc{R}$, $q$ is a predicate name not in $\mc{R}$, $\vec{X} \subseteq \bigcup
\vec{X}_i$ and the $\vec{X}_i$ are sequences of variables or constants. A
Boolean CQ (BCQ) over $\mc{R}$ is a CQ having head predicate $q$ of arity 0
(i.e., no variables in $\vec{X}$). The answer to a CQ $q(\vec{X}) \leftarrow
p_1(\vec{X}_1),\ldots,p_n(\vec{X}_n)$ over an instance $I$, denoted as $q(I)$,
is the set of all $n$-tuples $t \in \Gamma^n_C$ for which there exists a
homomorphism $h$ such that $h(p_1(\vec{X}_1),\ldots,p_n(\vec{X}_n)) \subseteq
I$ and $h(\vec{X}) = t$.  A BCQ
has only the empty tuple~$\tup{}$ as possible answer, in which case we say it
has a positive answer, denoted $I \models q$.

A program $\mc{P}$ consists of a set of rules $\Sigma^\mc{P}$ and a database
$D^\mc{P}$ over same schema $\mc{R}$.  $\mc{P}$ is a ground program if
$\Sigma^\mc{P}$ is a set of ground rules.  Given a program $\mc{P}$ and a CQ
$q$, the answers to $q$ are those that are true in all models of $\mc{P}$.
Formally, the models of $\mc{P}$, denoted as $\nit{mods}(\mc{P})$, is the set
of all instances $I$ such that $I \supseteq D^\mc{P}$ and $I \models
\Sigma^\mc{P}$.  The answers to a CQ $q$ over $\mc{P}$, denoted as
$\nit{ans}(q, \mc{P})$, is the set of n-tuples $\{t\;|\;t \in q(I), \forall I
\in \nit{mods}(\mc{P})\}$.  The answer to a BCQ $q$ is positive, denoted as
$\mc{P} \models q$, if $\nit{ans}(q, \mc{P})$ is not empty.

\vspace{-3mm}

\subsection{Chase and Grounding}\label{sec:chase}

\vspace{-1mm}
The \emph{chase} procedure is a fundamental algorithm in various database
problems including implication of database dependencies, query containment and
CQ answering under dependencies~\cite{beeri,johnson,maier}.
The chase has been broadly employed in CQ~answering in the presence of
dependencies~\cite{cali12,fagin}; the intuition is that, given a set of
dependencies over a database schema and a fixed database instance as input,
the chase ``repairs'' the instance so that the result satisfies the
constraints.
The result of the chase procedure, also called chase, is a so-called
\emph{universal model}~\cite{fagin}, i.e., a representative of all models in
$\nit{mods}(\mc{P})$; therefore, the answers to a CQ $q$ under dependencies
(in the open-world assumption, also called \emph{certain answers}), can be
computed by evaluating $q$ over the chase (and discarding the answers
containing labeled nulls).  The chase under TGDs, which we do not describe in
detail, is in fact a form of \emph{grounding}; in Section~\ref{sec:grounding}
we propose a grounding technique for answering CQs under WS~\datalogpm{} based
on a variant of the chase.

\vspace{-2mm}

\subsection{\datalogpm{} and the Stickiness Paradigm}
\vspace{-1mm}

Query answering with respect to a set of TGDs is generally undecidable as
proved in~\cite{beeri-chase-termination}. The \datalogpm{} family contains syntactic classes of
TGDs that impose restrictions on the form of the rules to guarantee
decidability and in many cases tractability of query answering. Two relevant
decidability paradigms are \emph{guardedness} and \emph{stickiness}.
In this paper we concentrate on stickiness, which is a syntactic condition on
the join variables in the body of the rules.

\vspace{-0.5cm}
\subsubsection{Sticky programs} \label{sec:sticky} Sticky rules are defined by
means of a body variable {\em marking procedure} that takes as input the set
$\Sigma^\mc{P}$ of rules. It has two steps:\vspace{-2mm}
\begin{enumerate}
\item \textsl{Preliminary step}: for each $\sigma \in \Sigma^\mc{P}$ and for
  each variable $X \in {\it body}(\sigma)$, if there is an atom $a \in {\it
    head}(\sigma)$ such that $X$ does not appear in $a$, mark each occurrence
  of $X$ in ${\it body}(\sigma)$.
\item \textsl{Propagation step}: for each $\sigma \in \Sigma^\mc{P}$, if a
  marked variable in ${\it body}(\sigma)$ appears at position $\pi$, then for
  every $\sigma' \in \Sigma^\mc{P}$ (including $\sigma$), mark each occurrence
  of the variables in ${\it body}(\sigma')$ that appear in ${\it
    head}(\sigma')$ in the same position $\pi$.
\end{enumerate}

\vspace{-3mm}
\begin{example} \label{example:sticky} Consider a program $\mc{P}$, with
  $\Sigma^\mc{P}$ as following set of rules in which the marked variables
  (denoted by hat signs) after applying the preliminary
  step: 
\vspace{-2mm}
\[
  \begin{array}{ccc}
    r(X,Y),p(X,Z) \rightarrow s(X,Y,Z). &\hspace*{1.2em}&
    u(X) \rightarrow \exists\;Y\;r(Y,X).
  \end{array}
\]
\vspace{-5mm}
\[
  \begin{array}{c}
    s(\hat{X},Y,\hat{Z}) \rightarrow u(Y).
  \end{array}
\]
  In the first rule, variables $X$ and $Z$ are marked after applying one
  propagation step since they appear in the head in marked positions ($s[1],s[3]$), and the final, marked rules are:
\vspace{-2mm}
\[
  \begin{array}{ccc}
    r(\hat{X},Y),p(\hat{X},\hat{Z}) \rightarrow s(X,Y,Z). &\hspace*{1.2em}&
    u(X) \rightarrow \exists\;Y\;r(Y,X).
  \end{array}
\]
\vspace{-5mm}
\[
  \begin{array}{c}
    s(\hat{X},Y,\hat{Z}) \rightarrow u(Y).
  \end{array} \vspace*{-1.7em}
\]
\boxtheorem \end{example}

\vspace*{-1em}
$\mc{P}$ is sticky when, at the end of the marking procedure over
$\Sigma^\mc{P}$, there is no rule with a marked variable in its body that
occurs more than once. From Example~\ref{example:sticky}, we can see that the
program is {\em not} sticky since $X$ in the first rule is marked and occurs
twice in $r[1]$ and $p[1]$.

Sticky programs enjoy \emph{first-order rewritability}. A class of programs is
first-order rewritable if, for every program $\mc{P}$ in the class, and for
every BCQ $q$, there is a first-order query $q_\mc{P}$ such that, $\mc{P}
\models q$ if and only if $D^\mc{P} \models q_\mc{P}$. Thus, under first-order
rewritable programs, CQs can be answered by constructing the (finite)
rewritten first-order query~\cite{gottlob14}, and then evaluating it over the
extensional database. Since evaluation of first-order queries is in
\textsc{ac}$_0$ in data complexity~\cite{abiteboul95}, it immediately follows
that CQ answering under first-order rewritable classes of rules, including
sticky ones, is in \textsc{ac}$_0$.

\vspace{-4mm}
\subsubsection{Weakly-Sticky programs} \label{sec:ws} Weakly-sticky (WS) programs
generalize sticky programs and the well known class of weakly-acyclic
programs. The class of WS programs is not first-order rewritable
but CQ answering over these programs is proved to be tractable in data
complexity~\cite{cali12}. The definition of WS programs appeals to
conditions on repeated variables in its rule bodies, and is based on the
notion of {\it dependency graph} and the positions with finite rank in such a
graph~\cite{fagin} that we explain in Example~\ref{example:dg}.

\vspace{-2mm}
\begin{example} \label{example:dg} (Example \ref{example:sticky} cont.)
  Consider the set of rules $\Sigma^\mc{P}$ in program $\mc{P}$. The
  dependency graph of $\Sigma^\mc{P}$ is a directed graph constructed as
  follows: The vertices in $V$ are positions of the predicates in schema of
  $\Sigma^\mc{P}$, and the edges in $E$ are defined as it follows. For every
  $\sigma \in \Sigma^\mc{P}$ and non-existential variable $x$ in ${\it
    head}(\sigma)$ and in position $\pi$ in ${\it body}(\sigma)$: (1) for each
  occurrence of $x$ in position $\pi'$ in ${\it head}(\sigma)$, create an edge
  from $\pi$ to $\pi'$; (2) for each existential variable $z$ in position $\pi''$ in
  ${\it head}(\sigma)$, create a \emph{special edge} from $\pi$ to $\pi''$.
  The dependency graph of $\Sigma^\mc{P}$ from Example~\ref{example:sticky} is
  illustrated in Figure~\ref{fig:dg}. The special edge from $u[1]$ to $r[1]$
  is shown by a dotted arrow that depicts values invention by existential
  variable $Y$ in the last rule.
  \begin{figure}[t]
    \begin{center}
      \vspace{-4mm}
       \resizebox{20em}{!}{\input{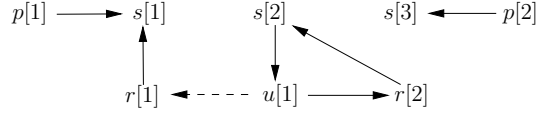}}
      \caption{Dependency graph for $\Sigma^\mc{P}$ in
        Example~\ref{example:dg}}\label{fig:dg}
      \vspace{-10mm}
    \end{center}
  \end{figure}
  The \emph{rank} of a position is the maximum number of special edges over
  all (finite or infinite) paths ending at that position. Accordingly,
  $\Pi_F(\Sigma^\mc{P})$ denotes the set of positions of finite rank, and
  $\Pi_\infty(\Sigma^\mc{P})$ the set of positions of infinite
  rank. Intuitively, $\Pi_F(\Sigma^\mc{P})$ captures positions where finitely
  many values may appear during the chase; and $\Pi_\infty(\Sigma^\mc{P})$
  those where infinitely many fresh null values may occur during the chase.
  In this example, $\Pi_\infty(\Sigma^\mc{P})$ is empty. The rank of $u[1],
  s[2], r[2], s[3], p[1]$ and $p[2]$ is zero and the rank of $r[1]$ and $s[1]$
  is one. \boxtheorem\end{example}

\vspace{-3mm}

A program $\mc{P}$ is WS if for every rule in $\Sigma^\mc{P}$ and every
variable in its body that occurs more that once, the variable is either
non-marked or appears at least once in a position in $\Pi_F(\Sigma^\mc{P})$
(position with finite rank).

\vspace{-2mm}

\begin{example} (Example \ref{example:dg} cont.) $\mc{P}$ is
  WS. $\Pi_F(\Sigma^\mc{P})$ contains $s[3]$, $p[1]$ and $p[2]$ and
  the other positions of the predicates in $\Sigma^\mc{P}$ are in
  $\Pi_\infty(\Sigma^\mc{P})$. The repeated variable $X$ in the body of the
  second rule is marked, but it appears at least once in the finite-rank
  position $p[1]$.\boxtheorem
\end{example}


\vspace{-5mm}

\section{Grounding Based on a Query-Driven Chase}
\label{sec:grounding}

\vspace{-2mm}

In this section we adopt the chase procedure proposed in~\cite{milani15-amw}
for query answering under WS \datalogpm{} programs as a basis for a
query-driven grounding algorithm, called {\sf GroundWS}. {\sf GroundWS} takes as input a WS \datalogpm{} program $\mc{P}$ and a CQ $q$
and returns a ground program $\mc{P}'$ for which $q$ can be efficiently
answered.

The {\sf GroundWS} algorithm uses the notion of applicability that we explain
next. A rule $\sigma$ and homomorphism $h$ are \emph{applicable} if: (a)
$h({\it body}(\sigma)) \subseteq\mc{H}\cup D^\mc{P}$ in which $\mc{H}$ is the set
of head atoms of $\Sigma^{\mc{P}'}$ (the set of already grounded rules); (b)
There is no atom $a \in \mc{H}$ and homomorphism $h'$ as an extension of $h$
such that $h'$ maps existential variables in $\sigma$ to fresh nulls in
$\Gamma_N$ and $h'({\it head}(\sigma))$ is $\Pi_F(\Sigma^\mc{P})$-homomorphic
to $a$.
The second applicability condition is imposed to prevent infinite grounding
steps as we describe next. Importantly, we use
$\Pi_F(\Sigma^\mc{P})$-homomorphism instead of ordinary homomorphism to
consider only the null values that appear in the positions with infinite rank,
which is sufficient to prevent infinite grounding steps.
We now illustrate the algorithm \textsf{GroundWS}, which consists of the
following steps:
\begin{enumerate}
\item Initialize $\Sigma^{\mc{P}'}$ to $\emptyset$ and $D^{\mc{P}'}$ to
  $D^\mc{P}$.
\item For every rule $\sigma \in \Sigma^\mc{P}$ applicable with homomorphism
  $h$, add $h'(\sigma)$ into $\Sigma^{\mc{P}'}$, in which $h'$ is extension of
  $h$ that maps existential variables of $\sigma$ into fresh null values in
  $\Gamma_N$.
\item Apply Step~2 iteratively, until there are no more applicable rules.
\item Resume Step~2 after freezing every labeled null value in
  $\Sigma^{\mc{P}'}$, where by freezing a null we mean replacing it with a
  special constant in $\Gamma_C$, which henceforth is considered as a constant
  but is never returned in the result of a query.  Repeat this resumption
  $M_q$ times, where $M_q$ is the number of variables in $q$.
\item Return $\mc{P}'$.
\end{enumerate}

Notice that every pair of rule and homomorphism in Step~2 is applied only
once. Moreover, if there are more than one pairs of applicable rule\textbackslash homomorphism,
then {\sf GroundWS} applies them in a level saturating fashion. More
specifically, {\sf GroundWS} chooses the rule and the homomorphism for which
the body atoms have the smallest maximum level.  Here the level of an atom
is $0$ if it is in $D^\mc{P}$, and it is the maximum level of the body atoms
of a rule plus one for the head atom of the rule.
%
It is important to notice that by freezing a null value we consider it as a
constant only for deciding homomorphic atoms (specifically in the second
applicability condition), and not during query answering. That is, the frozen
nulls still can not appear in query answers since they are not in the active
domain of the extensional database.
%

\begin{example} Consider WS program $\mc{P}$ with $D^\mc{P}= \{p(a,b),c(b)\}$,
  and a BCQ $q \leftarrow u(X)$, and a set of rules $\Sigma^\mc{P}$ as
  follows:
  \[
  \begin{array}{rl c rl}
    \sigma_1\!:\;\;\; & p(X,Y)~\rightarrow~\exists Z\;p(Y,Z). &\hspace{1cm}&
    \sigma_2\!:\;\;\;& p(X,Y),c(X),p(Y,Z)~\rightarrow~u(Y).
  \end{array}
  \]


  We start from $D^\mc{P}$ and iteratively generate ground rules by mapping
  via homomorphism the body of the rules in $\Sigma^\mc{P}$ into $D^\mc{P}$ or
  the head of the rules in $\Sigma^{\mc{P}'}$ (the current set of ground
  rules).  The basic algorithm is as follows: we iteratively add a ground rule
  to $\Sigma^{\mc{P}'}$ if its head atom is not homomorphic to the head of a
  rule already in $\Sigma^{\mc{P}'}$, until no new rule can be added.  This
  ``cautious'' procedure, similar to the chase procedure
  in~\cite{leone11,milani15-amw}, guarantees termination.
  In our example, {\sf GroundWS} stops after adding only one ground rule
  $\sigma'_1:p(a,b)~\rightarrow~p(b,\zeta_1)$ to $\Sigma^{\mc{P}'}$, where
  $\zeta_1$ is a labelled null in $\Gamma_N$.
  In order to complete our grounding, we \emph{resume} the above basic
  algorithm $M_q$~times, where $M_q$ is the number of variables in
  $q$.  Before each resumption, we \emph{freeze} the labelled nulls, which
  after having frozen are considered as constants.  In our example, we have
  only another resumption, which adds the rules
  $\sigma''_1:p(b,\zeta_1)~\rightarrow~p(\zeta_1,\zeta_2)$ and $\sigma'_2:\,
  p(b,\zeta_1),c(b),p(\zeta_1, \zeta_2)~\rightarrow~u(\zeta_1)$.  Resumptions
  are needed, intuitively, to capture applications of rules in the chase
  procedure where a join variable, appearing in two or more distinct atoms, is
  mapped to a labelled null.\boxtheorem \end{example}

Notice that the number of resumptions depends on the query, which makes our
grounding also dependent of the query; however, for practical purposes, we
could ground with $N$~resumptions so as to be able to answer queries with up
to ~$N$ existential variables, and if a query with more than $N$ existential
variables is to be answered, we can incrementally retake the already-computed
grounding and add the required number of resumptions.

\begin{theorem}\label{th:grounding} \em For every WS program $\mc{P}$ and CQ
  $q$, ${\sf GroundWS}$ runs in \textsc{ptime} with respect to the size of
  $D^\mc{P}$ and in \textsc{2exptime} with respect to the size of $\mc{P}$ and $q$, and returns a ground program $\mc{P}'$ such that,
  $\nit{ans}(q,\mc{P})=\nit{ans}(q,\mc{P}')$.
\end{theorem}

\noindent \textsl{Proof (sketch):} ${\sf GroundWS}$ runs in polynomial time
essentially because of the second applicability condition, which at each
resumption prevents the generation in in $\Sigma^{\mc{P}'}$ of two distinct
(ground) rules having $\Pi_F(\Sigma)$-homomorphic heads;
since the number of terms in the positions of $\Pi_F(\Sigma)$ is polynomial
with respect to $D^\mc{P}$~\cite[Theorem 3.9]{fagin}, the~\textsc{ptime}
membership follows.  Notice that the above condition, within a
resumption ``phase'', prevents the generation of some (ground) rules in
$\Sigma^{\mc{P}'}$ that are necessary for answering $q$; these rules depend on
replacing a join variable with a null value in the rule body; the subsequent
resumption phase adds at least one of such rules to $\Sigma^{\mc{P}'}$. The~\textsc{2exptime} combined complexity is because exponentially many terms can appear in the positions of $\Pi_F(\Sigma)$ with respect to the size of $\mc{P}$ and $q$. This is implicit in the proof of~\cite[Theorem 3.9]{fagin}. As a result, the total number of atoms in each resumption phase is double exponential in the size of $\mc{P}$ and $q$.

The weak-stickiness of $\Sigma^{\mc{P}}$ implies that such null
values continue to appear in the head of consequent ground rules all the way
to the head atoms mapped to the query.  Hence there are at most $M_q$ such
rules generated for answering $q$, given that each rule ``saturates'' one of
the $M_q$ existential variables in $q$.  Therefore only $M_q$ resumptions are
necessary.
\wboxtheorem

\section{Partial Grounding for Weakly-Sticky \datalogpm{}}
\label{sec:partial}

In this section we propose a partial grounding algorithm, called {\sf
  PartialGroundingWS}, that takes a WS \datalogpm{} program $\mc{P}$ and
transforms it into a sticky \datalogpm{} program $\mc{P}'$ such that $\mc{P}'$
is equivalent to $\mc{P}$ for CQ answering. \textsf{PartialGroundingWS}
selectively replaces certain variables in positions of finite rank with
constants from the active domain of the underlying database.
Our algorithm requires that the set of rules $\Sigma^\mc{P}$ in the input
program satisfies the condition that there is no existential variable in
$\Sigma^\mc{P}$ in any finite-rank position; therefore each position in the
input set of rules~$\Sigma^\mc{P}$ will have rank either~$0$ or~$\infty$.  The
reason for this requirement is the convenience of grounding variables at
zero-rank positions by replacing them by constants rather than by
labeled nulls. This does not really restrict the input programs since, as we will show later, an arbitrary program can be transformed by the {\sf ReduceRank} algorithm to a program that has the requirement.

%
Before illustrating the \textsf{PartialGroundingWS} algorithm, we therefore
present an algorithm, called \textsf{ReduceRank}, that takes a program
$\mc{P}$ and compiles it into an equivalent program $\mc{P}'$ whose rule set
$\Sigma^{\mc{P}'}$ has only zero-rank or infinite-rank positions.
%
The \textsf{ReduceRank} algorithm is inspired by the reduction method
in~\cite{krotzsch} for transforming a weakly-acyclic program into an
existential-free Datalog program. Given a program $\mc{P}$,
\textsf{ReduceRank} executes of the following steps.

\begin{enumerate}
\item Initialize $\Sigma^{\mc{P}'}$ to $\Sigma^\mc{P}$ and $D^{\mc{P}'}$ to
  $D^\mc{P}$.
\item Choose a rule $\sigma$ in $\Sigma^{\mc{P}'}$ with an existential
  variable in a position with rank~1.  Notice that if there are existential
  variables in positions with finite rank, at least one of the positions has
  rank~1.
\item Skolemize $\sigma$ as $\sigma'$ by replacing the existential variable
  with a functional term.  For example, $\sigma:p(X,Y)\rightarrow\exists
  Z\;r(Y,Z)$, becomes $\sigma':p(X,Y)\rightarrow r(X,f(X))$.
\item Replace the Skolemized predicate $r$ that has the function term with a
  new expanded predicate of higher arity (the arity of~$r$ plus~$1$) and
  introduce a \emph{fresh} special constant of~$\Gamma_C$ to represent the
  function symbol.  The new constant precedes its arguments in a newly
  introduced position.  For example, $r(X,f(X))$ becomes,
  $r'(X,\mathsf{f},X)$; we are therefore expanding the position~$r[2]$.
  \ignore{\begin{noteac}
    Sorry but I forgot the details here.  If we have $f(X,g(Y))$ the arguments
    are $\tup{\mathsf{f},X,\mathsf{g},Y}$; how do we know the correct nesting
    of the terms? How do we know that $\mathsf{g}$ is within the second
    argument of~$\mathsf{f}$?
  \end{noteac}}
\item Replace the expanded predicate in other rules and analogously expand
  other predicates in positions where variables appearing in the expanded
  position appear: if a variable appears in $\sigma \in \Sigma^\mc{P}$ in an
  expanded position $\pi$ and also in another position $\pi_1$, then
  also~$\pi_1$ is expanded (with its predicate) and the same variables of
  $\pi$ also appear in $\pi_1$, thus preserving the join.  More precisely, let
  $X$ be a variable at position~$\pi$ in an atom~$r(\ldots,X,\ldots)$ that is
  expanded in some rule $\sigma$ into $r(\ldots,X_1,X_2,\ldots)$; if there is
  another rule $\sigma_1 \in \Sigma^\mc{P}$ containing atoms $r(\ldots,Y,\ldots)$ and
  $s(\ldots,Y,\ldots)$ such that $Y$ appears in the former at position~$\pi_1$
  and in the latter at position~$\pi_2$, then the first atom is replaced by
  $r'(\ldots,Y_1,Y_2,\ldots)$ (expanded version on~$\pi_1$ with the new
  variable~$Y$) and the second one is replaced by $s'(\ldots,Y_1,Y_2,\ldots)$
  (expanded version on~$\pi_2$).
  \ignore{\begin{noteac}
    I am not sure if what I wrote above is fully correct; I tried to be more
    precise.  Mostafa, please have a look.
  \end{noteac}}
For example, $r(X,Y),t(Y,Z) \rightarrow s(X,Y,Z)$ becomes
$r'(X,Y,Y'),t'(Y,Y',Z) \rightarrow s'(X,Y,Y',Z)$. Notice that if a predicate
is expanded in a head-atom in a position where an existential variable occurs,
the new positions are not required and are filled with the special symbol
${\small \triangle}$.
  \ignore{\begin{noteac}
    The use of the filler is unclear here.  We need a proper definition and a
    proper example.  Mostafa, can you do something about it?
  \end{noteac}}
\item If the expanded predicates have extensional data, add new rules to
  $\Sigma^{\mc{P}'}$ to ``load'' the extensional data into the expanded
  predicates. For example, if $r$ has extensional data, we add a rule, $r(X,Y)
  \rightarrow r'(X,Y,{\small \triangle})$. Here, ${\small \triangle}$ is used to fill the new position in the expanded predicate since it does not carry extensional data.
  \ignore{\begin{noteac}
    Here the example isn't fully clear; is the filler~$\Box$ (maybe we
    should use another symbol, e.g.~$\triangle$) just a filler to render its
    occurrence irrelevant?  Does it join with other fillers (I assume it
    doesn't, but we should maybe make this clear).
  \end{noteac}}
\item Repeat Steps~$2$ to~$6$ until there is no existential variable in a
  finite-rank position.
\end{enumerate}

Note that in Step~3 only the body variables that also appear in the head
participate as arguments of the function term.  For example, in the
Skolemization of $p(X,Y)\rightarrow\exists Z\;r(Y,Z)$, the function term does
not include~$X$ since the rule can be broken down into $p(X,Y)\rightarrow
u(Y)$ and $u(Y)\rightarrow \exists Z\;r(Y,Z)$. Notice that given a CQ $q$ over $\mc{P}$, Steps 2 to 6 are also applied on $q$ obtaining a new CQ $q'$ over $\mc{P}'$.


\begin{example} Let $\mc{P}$ be a program with $\Sigma^\mc{P}$ as follows.
\[
  \begin{array}{rlcrl}
    \sigma_1:\;\;& v(X)~\rightarrow~\exists Y\;r(X,Y). &\hspace{1cm}&
    \sigma_3:\;\; & t(X,Y),v(X)~\rightarrow~ p(X,Y).\\
    \sigma_2:\;\; & r(X,Y)~\rightarrow~ \exists Z\;t(X,Z). &\hspace{1cm}&
    \sigma_4:\;\; & p(X,Y)~\rightarrow~\exists Z\;p(Y,Z).
  \end{array}
\]
  In this program, $\Pi_F(\Sigma^\mc{P})=\{v[1], r[1], r[2], t[1],
  t[2]\}$. {\sf ReduceRank} will eliminate $Y$ in $\sigma_1$ and $Z$ in
  $\sigma_2$, but not $Z$ in $\sigma_4$ since the later is in an infinite rank
  position. {\sf ReduceRank} chooses $Y$ in $\sigma_1$ over $Z$ in $\sigma_2$
  since, $Y$ is in $r[2]$ with zero rank and $Z$ is in $t[2]$ with rank
  1. After applying Steps~2-6, $\sigma_1$ and $\sigma_2$ become
  $\sigma'_1:~v(X)~\rightarrow~r'(X,{\sf f},X)$ and
  $\sigma'_2:~r'(X,Y,Y')~\rightarrow~ \exists Z\;t(X,Z).$ By removing $Y$ from
  $\sigma_1$, $Z$ in $\sigma_2$ is placed in a position with zero rank. {\sf
    ReduceRank} repeats Steps~2-6 to eliminate $Z$ in $\sigma_2$ which results
  into $\Sigma^{\mc{P}'}$:
\[
  \begin{array}{c c c}
    v(X)\rightarrow r'(X,{\sf f},X).&\hspace{5mm}& r'(X,Y,Y')\rightarrow t'(X,{\sf g},X).\\
    t'(X,Y,Y'),v(X) \rightarrow p'(X,{\small \triangle},Y,Y').&\hspace{5mm}& p'(X,X',Y,Y') \rightarrow \exists Z\;p'(Y,Y',Z,{\small \triangle}).\\
  \end{array}
\]
  Notice that {\sf ReduceRank} does not try to remove $Z$ in the last rule,
  since it is in the infinite rank position $p[3]$. Note also that $p$ is
  expanded twice since both its positions can host labeled nulls generated by
  $Z$ in $\sigma_2$. \boxtheorem
\end{example}

\begin{proposition}\label{prop:complexity}\em Given a CQ $q$ over a program $\mc{P}$, {\sf
    ReduceRank} runs in \textsc{exptime} with respect to the size of
  $\Sigma^\mc{P}$ and returns a CQ $q'$ over a program $\mc{P}'$ such that $\mc{P}'$ has no existential
  variable in finite rank positions of $\Sigma^{\mc{P}'}$ and $\nit{ans}(q,\mc{P})=\nit{ans}(q',\mc{P}')$.
\end{proposition}

For every rule in $\Sigma^\mc{P}$, there is only one corresponding rule in
$\Sigma^{\mc{P}'}$. There are also rules in $\Sigma^{\mc{P}'}$ for loading the
extensional data of the expanded predicates. Therefore, the number of rules in
$\Sigma^\mc{P}$ is the same order of the size of $\Sigma^{\mc{P}'}$. The arity of the predicates in $\Sigma^{\mc{P}'}$ can have an exponential increase with respect to the arity of predicates in $\Sigma^\mc{P}$ which makes {\sf ReduceRank} run in
\textsc{exptime}. $\mc{P}$ and $\mc{P}'$ are equivalent since the expanded
predicate that represent the propagation of null values in $\mc{P}$ are
applied on every possible rule in $\Sigma^{\mc{P}'}$ in Steps~5-6 .

Notice that \textsf{ReduceRank} preserves the weak-stickiness property because
the property only concerns repeated marked variables that occur in infinite
rank positions, while {\sf ReduceRank} involves finite rank positions and it
does not create a new marked variable or a new infinite rank position to break
the property.  We can therefore state the following.

\begin{lemma}\label{lem:ws} \em The class of WS programs is closed under {\sf
    ReduceRank}.
\end{lemma}


Now that we explained the {\sf ReduceRank} algorithm, we continue and
present the {\sf PartialGroundingWS} algorithm. Given a WS program $\mc{P}$,
let us call \emph{weak rules} the rules of $\Sigma^\mc{P}$ in which some
repeated marked body variables (which we call weak variables) appear at least once in a position with finite
rank.  {\sf PartialGroundingWS}
transforms $\mc{P}$ into a sticky program $\mc{P}'$. The sticky program
$\mc{P}'$ has the same database as $\mc{P}$ ($D^{\mc{P}'}=D^\mc{P}$) and its
set of rules $\Sigma^{\mc{P}'}$ is obtained by replacing the weak variables of
$\Sigma^\mc{P}$ with every constants from the active domain of
$D^\mc{P}$. Example~\ref{exp:partial} illustrates the {\sf PartialGroundingWS}
algorithm.

\begin{example} \label{exp:partial} Consider a WS program $\mc{P}$ with
  $D^\mc{P}=\{p(a,b),r(a,b)\}$ and $\Sigma^\mc{P}$ consisting of the following
  rules:
\[
  \begin{array}{rlcrl}
    \sigma_1\!: & p(\hat{X},\hat{Y})~\rightarrow~\exists Z\;p(Y,Z).&\hspace{1cm}&
    \sigma_3\!: & s(\hat{X},Y,Z),r(\hat{X},Y) \rightarrow t(Y,Z).
  \end{array}
\]
\[
  \begin{array}{rl}
    \sigma_2\!: & p(\hat{X},Y),p(Y,Z)~\rightarrow~s(X,Y,Z).
  \end{array}
\]
  Here $\sigma_3$ is a weak rule with $X$ as its weak variable.  Notice that
  $Y$ in $\sigma_2$ and $\sigma_3$ are not weak since they are not marked
  (the hat signs show the marked variables). We replace $X$ with constants $a$ and $b$
  from $D^\mc{P}$. The result is a set of sticky rules $\Sigma^{\mc{P}'}$ that
  includes $\sigma_1$ and $\sigma_2$ as well as the following rules,
  $\sigma'_3:~s(a,Y,Z),r(a,Y) \rightarrow t(Y,Z)$ and
  $\sigma''_3:~s(b,Y,Z),r(b,Y) \rightarrow t(Y,Z)$.\boxtheorem
\end{example}

\vspace{-3mm}
\begin{theorem}\label{th:partial} \em Let $\mc{P}$ be a WS program such that
  there is no existential variable in $\Sigma^\mc{P}$ in a finite rank
  position. {\sf PartialGroundingWS} runs in polynomial time with respect to
  the size of $D^\mc{P}$ and it transforms $\mc{P}$ into a sticky program
  $\mc{P}'$ such that for every CQ $q$, the following holds:
  $\nit{ans}(q,\mc{P})=\nit{ans}(q,\mc{P}')$.
\end{theorem}

\noindent \textsl{Proof (sketch):} $\mc{P}'$ is sticky since every weak
variable, that by its definition breaks the weak-stickiness, is
grounded. $\mc{P}$ and $\mc{P}'$ are equivalent for query answering since the
weak variables of $\Sigma^\mc{P}$ are replaced in $\Sigma^{\mc{P}'}$ with
every possible constant from $D^\mc{P}$, and based on our assumption on
$\Sigma^\mc{P}$, only constants can substitute these variables. Additionally,
{\sf PartialGroundingWS} runs in polynomial time since weak variables are
replaced with constants from $D^\mc{P}$.\wboxtheorem

A possible optimization for {\sf PartialGroundingWS} is to narrow down the
values for replacing the weak variables, that is to ignore those constants in
the active domain of $D^\mc{P}$ that can not appear in the positions where
weak variables appear during the chase of~$\mc{P}$.  In
Example~\ref{exp:partial}, $\sigma'_3$ is not useful since $a$ can never be
assigned to $X$ in $\sigma_3$.
For this purpose, {\sf GroundWS} can be applied to compute the possible values
for partial grounding.  For example, a CQ,
$s(X,Y,Z),r(X,Y)~\rightarrow~q_g(X)$ returns constants for grounding the weak
variable $X$ in $\sigma_3$.


\vspace{-0.2cm}
\section{A Hybrid Approach}
\label{sec:hybrid}

In this section we propose a query answering algorithm for WS programs based
on a hybrid approach that combines {\sf ReduceRank} and {\sf
  PartialGroundingWS} from the previous section with a query rewriting
algorithm for sticky programs~\cite{gottlob14}.
Given a WS program $\mc{P}$ and a CQ $q$, hybrid query answering proceeds as
follows:
\begin{enumerate}
\item Use {\sf ReduceRank} to compile $\mc{P}$ into a WS program $\mc{P}'$
  with no existential variable in finite rank positions.
\item Apply {\sf PartialGroundingWS} on $\mc{P}'$ that results to a sticky
  program $\mc{P}''$.
\item Rewrite $q$ into a first-order query $q'$ using the rewriting algorithm
  proposed in~\cite{gottlob14} and answer $q'$ over $D^{\mc{P}''}$ (any other sound and complete rewriting algorithm for sticky programs is also applicable at this step).
\end{enumerate}

\begin{example}\label{exp:final} Consider a WS program $\mc{P}$ with database
  $D=\{v(a)\}$ and $\Sigma^\mc{P}$ consisting of the following rules:
  \[
\arraycolsep=15pt
  \begin{array}{rlcrl}
    \sigma_1: &\hspace{-5mm}p(X,Y)~\rightarrow~\exists
    Z\;p(Y,Z). &\hspace{-15mm}& \sigma_4: &\hspace{-5mm}r(X,Y),s(X,Z) \rightarrow c(Z).\\
    \sigma_2: &\hspace{-5mm}p(X,Y),p(Y,Z)~\rightarrow~u(Y). &\hspace{-15mm}& \sigma_5: &\hspace{-5mm}c(X)~\rightarrow~\exists Y\;p(X,Y).
  \end{array}
  \]
  \vspace{-5mm}
  \[
  \begin{array}{rl}
    \sigma_3: \;\;\;&\hspace{-5mm}v(X)\rightarrow~\exists Y\;r(X,Y).
  \end{array}
  \]

  The {\sf ReduceRank} method removes the existential variable $Y$ in
  $\sigma_3$. The result is a WS program $\mc{P}'$ with $\Sigma^{\mc{P}'}$:
  \[ \arraycolsep=15pt
  \begin{array}{cc}
    p(X,Y)~\rightarrow~\exists Z\;p(Y,Z). & r'(X,Y,Y'),s(X,Z) ~\rightarrow~ c(Z).\\
    p(X,Y),p(Y,Z)~\rightarrow~u(Y). & c(X)~\rightarrow~\exists Y\;p(X,Y).
  \end{array}
  \]
  \vspace{-5mm}
  \[
  \begin{array}{c}
    v(X)~\rightarrow~ r'(X,{\sf f},X).
  \end{array}
  \]


  Next, {\sf PartialGroundingWS} grounds the only weak variable, $X$ in
  $\sigma'_4$ with constant $a$ which results into sticky program $\mc{P}''$
  with
  $\Sigma^{\mc{P}''}=\{\sigma_1,\sigma_2,\sigma'_3,\sigma''_4,\sigma_5\}$, in
  which $\sigma''_4:~r'(a,Y,Y'),s(a,Z) \rightarrow c(Z)$. $\mc{P}''$ is sticky and a CQs can be answered by rewriting it in terms of
  $\Sigma^{\mc{P}''}$ and answered directly on
  $D^{\mc{P}''}=D^{\mc{P}}$. \boxtheorem
\end{example}


\begin{corollary} \em Given a WS program $\mc{P}$ and a CQ $q$, the set of
  answers obtained from the hybrid approach is $\nit{ans}(q,
  \mc{P})$.
\end{corollary}






\vspace{-0.5cm}
\section{Conclusions}
\label{sec:conclusion}


WS \datalogpm{} is an expressive ontology language with good
computational properties and capable of capturing the most prominent Semantic
Web languages.  We proposed two deterministic algorithms for answering
conjunctive queries on WS \datalogpm.  In the first algorithm, a
variant of the well-known chase, which proceeds in terminating
``resumptions'', generates an expansion of the given database that contains
all (ground) atoms needed to answer the query; the expansion depends on the
query as the number of resumptions is the number of existentially quantified
variables of the query.  For practical purposes, one can expand up to~$m$
resumptions off-line, and the expansion will serve to answer all queries with
up to~$m$ existential variables.  If at a certain point a query with more
than~$m$ existential variables is to be processed, more resumptions can be
performed from the expansion already computed.  This, of course, if there are
no changes in the given database.  The second algorithm transforms a
WS program into a sticky one by means of \textsl{(a)} a
Skolemization and annotation procedure, which turns all finite-rank positions
into zero-rank ones, followed by \textsl{(b)} a partial grounding on the
zero-ranked positions.  Then, the rewriting technique for WS~programs is
employed; the rewriting, which is in the language of union of conjunctive
queries, is then evaluated directly on the given database.

\paragraph{Efficiency.}  Both algorithms we propose achieve the optimal lower
bound in data complexity (i.e., in complexity calculated having only the
database as input) for CQ answering under WS \datalogpm, that
is~\textsc{ptime}.  In the first algorithm, the expansion is computed
\emph{off-line}, and the final query processing step is a simple evaluation of
a CQ on an instance.  In the second algorithm, the rewriting is intensional
(i.e., it does involve the data) and the final step is the evaluation of a
union of CQs on the given database, which can easily done, for example, by
evaluating an SQL query on the database.

In the light of the above considerations, we believe that our contribution
sets the basis for practical query answering algorithms in real-world
scenarios.  We plan to continue our work by running experiments on large data
sets.  We also intend to refine the hybrid algorithm by limiting the number of
CQs in the final rewriting; to do so, we will avoid the grounding of rules
when we discover that having certain constants in certain position will not
yeld any new atom; such discovery can be performed by analyzing the dependency
graph and the TGDs in general.  This refinement will improve the efficiency of
the algorithm.





\appendix

\section{Appendix}
\label{sec:proofs}


\noindent {\bf Proof of Theorem~\ref{th:grounding}:} {{\sf GroundWS} runs in {\sc ptime} because of the second condition (Condition (b)) for applicable rule-homomorphism that is used in Step~2 of the algorithm. Specifically, the condition prevents from $\Sigma'$ multiple ground rules with $\Pi_F(\Sigma)$-homomorphic head atoms. Note that there are polynomially many values in the positions of $\Pi_F(\Sigma)$ during the algorithm. This follows from Theorem 3.9 in~\cite{fagin} that shows during the chase of a program $\mc{P}$, there are polynomially many values in the positions of $\Pi_F(\Sigma^\mc{P})$ with respect to the size of $D^\mc{P}$. As a result, the number of possible non-$\Pi_F(\Sigma^\mc{P})$-homomorphic atoms and also the number of possible ground rules at each resumption are polynomial with respect to the size of $D^\mc{P}$. Since the number of resumptions $M_q$ is independent of the data, the total number of ground rules in {\sf GroundWS} is also polynomial with respect to the size of $D^\mc{P}$ which proves the tractability of {\sf GroundWS}.

The second condition for applicable rule-homomorphism may prevent some ground rules from $\Sigma'$ that are necessary for answering $q$. Specifically, let $\sigma$ and $h$ be a rule-homomorphism that is not applicable because $b=h'(\nit{head}(\sigma))$ (with $h'$ extending $h$ with mapping for $\exists$-variables in $\sigma$) is $\Pi_F(\Sigma^\mc{P})$-homomorphic to an atom $a=\nit{head}(\sigma')$ with $\sigma' \in \Sigma'$. That means $a$ and $b$ differ in some null values, e.g. $\zeta_a$ and $\zeta_b$ resp that appear in non-$\Pi_F(\Sigma^\mc{P})$ positions. The null value $\zeta_b$ might appear in the head of the other ground rules in $\Sigma'$ and by preventing $h'(\sigma)$ (with the head $b$ that contains $\zeta_b$) from $\Sigma'$, we miss the query answers that depend (directly in the query's body or indirectly by the other applicable rules) on replacing a join variable with $\zeta_b$.

In order to solve this issue, we use resumptions at Step~4. More specifically, a resumption freezes $\zeta_a$ and $\zeta_b$ mentioned above by replacing them with fresh constants everywhere in $\Sigma'$. Consequently, any join that depends on replacing a join variable with $\zeta_b$ is enabled by augmenting $\Sigma'$ with $h'(\sigma)$. These joins might be in the body of the query or in the body of rules. In both cases the number of such joins is limited by $M_q$, the number of existential variables in the query. That is because of the weak-stickiness of $\mc{P}$ since the values that replace the join variables in non-$\Pi_F(\Sigma^\mc{P})$ positions continue to appear in the consequent head atoms and specifically those head atoms that are mapped to the body of the query. Therefore, after $M_q$ resumptions $\Sigma'$ contains every ground rule that is needed for answering $q$.\boxtheorem}


\noindent {\bf Proof of Proposition~\ref{prop:complexity}:} To prove that {{\sf ReduceRank} runs in {\sc exptime} with respect to the size of $\Sigma^\mc{P}$ we show that each iteration of Steps~2-6 runs in {\sc exptime}. Let consider one such iteration that transforms $\mc{P}_i$ into $\mc{P}_{i+1}$ and removes an existential variable $Y_i$ in the rule $\sigma_i$ and expands a predicate $p_i$ in the head of $\sigma_i$ to $p'_i$. Then, $\mc{P}_1$ is the input program $\mc{P}$ and $\mc{P}_n$ is the result program $\mc{P}'$ in which $n$ is the total number of iterations that is equal to the total number of existential variables in positions with finite rank.

$Y_i$ is in a position with the rank 1 and so expanding $p_i$ does not expand any predicate in the body of $\sigma_i$ during Step~5. As a result, every position in the program only gets expanded once during an iteration. Let $r$ and $b$ the number of rules and the maximum number of body atoms in $\Sigma^\mc{P}$ resp. that are not changed by the iterations. Also, let $w_i$ be and the maximum arity of atoms in $\Sigma^\mc{P}$ at the $i$-th iteration. Then, the arity of $p'_i$ (the expanded predicate of $p_i$) is at most increased by $b\times w_i$ (the maximum possible number of variables in the body of $\sigma_i$). Therefore, after propagating the expanded position in other rules, the maximum arity of the predicates $w_{i+1}=b\times w^2_i$ (each of the $w_i$ positions is expanded once by $b\times w_i$). This means after $n$ iterations the maximum arity of the predicates is $b^n\times w_i^{2n}$. On the other hand, $n$ is at most $w_1\times r$ and as the result $b^n\times w_i^{2n}$ is exponential with respect to $w_1$ and $r$ and so the size of $\Sigma^\mc{P}$ which proves our claim.

In order to prove the correctness of the transformation, i.e. $\nit{ans}(q,\mc{P})=\nit{ans}(q',\mc{P}')$, we prove the correctness of each iteration, $\nit{ans}(q_i,\mc{P}_i)=\nit{ans}(q_{i+1},\mc{P}_{i+1})$. We do that by constructing a model $I_{i+1}\models \mc{P}_{i+1}\cup q_{i+1}$ for every model $I_i \models \mc{P}^{i}\cup q_i$. Let assume that removing $Z_i$ introduces a function symbol {\sf f}. We first build an assignment $\mu_i$ that maps a null value in $I_i$ into a list of terms also in $I_i$. If $\sigma_i$ is the rule containing $Z_i$, for every homomorphism $h_i$ that maps both the body and the head of $\sigma_i$ into $I_i$, $\mu$ maps $h_i(Z_i)$ into the list of terms that appear in $h_i(\nit{body}(\sigma_i))$.

Now for every atom $a=p(t_1,...,t_n) \in I_i$ we add an atom $a'$ into $I_{i+1}$ that is constructed as follows: (a) if $p$ is not expanded in $\mc{P}_i$ then $a'=a$, (b) if $p$ is expanded into $p'$ in its $k$-th position, there are two possibilities, $t_k$ is either a null value or a constant. If $t_k$ is a constant expand it into $t_k,\triangle,...,\triangle$ to fill the expanded positions, and if $t_k$ is a null value, expand $t_k$ into $\mu(t_k)$.

$I_{i+1}$ is a model of $\mc{P}_{i+1}\cup q_{i+1}$ because for every homomorphism $g$ that maps the body of a rule $\sigma'$ into $I_{i+1}$ we can make an extension of $g'$ using $\mu$ that maps the head also into $I_{i+1}$.

In the other direction for any model $I_{i+1}$ of $\mc{P}_{i+1}\cup q_{i+1}$, we build a model $I_{i}$ of $\mc{P}_{i}\cup q_{i}$. This is simply by replacing any extended predicate $p'$ with its original predicate $p$ and removing additional terms, i.e. removing $\triangle$ symbols and the function symbols and their consequent terms with null values. For example, $p'(a,b,{\sf f},c)$ becomes $p(a,b,\zeta)$ if $p$ expanded by one, and $p'(a,b,d,\triangle)$ becomes $p(a,b,d)$. Again, it is straightforward to prove that $I_i$ is a model of $\mc{P}_i\cup q_i$ by showing that for any homomorphism that maps the body of a rule into $I_i$ there is an extension of it that maps the head into $I_i$.\boxtheorem}

\vspace{0.5cm}
\noindent {\bf Proof of Lemma~\ref{lem:ws}:} {To prove the class of weakly-sticky programs is closed under {\sf ReuceRank}, we show it is closed under each iteration of {\sf ReuceRank}, i.e. Steps~2-6, that removes one existential variable in a finite rank position. Let $\mc{P}_i$ be the inputs of the iteration and $\mc{P}_{i+1}$ be the result of the iteration. Now, we prove two claims about $\mc{P}_{i+1}$:

\begin{enumerate}
  \item Let $\pi$ be a position in $\mc{P}_{i}$ with finite rank and $\pi'$ the same position or one of the positions resulted by expanding $\pi$ in $\mc{P}_{i+1}$. If $\pi$ has finite rank, $\pi'$ also has finite rank. Using proof by contradiction, assume $\pi'$ is not finite, then there is a cycle in the dependency graph of $\mc{P}_{i+1}$ that includes $\pi$ and passes a special edge. The same positions in the cycle or their corresponding positions in $\mc{P}_{i}$, if they are expanded, make a cycle in the dependency graph of $\mc{P}_{i}$ that includes $\pi$ and also passes a special edge. Therefore, $\pi$ is not finite which contradicts the assumption and proves the claim.
  \item If a body variable $X$ in $\mc{P}^{i}$ is not marked, the same variable (or the variables $X_1,...,X_n$ if the position of $X$ is expanded in $\mc{P}_{i+1}$) is not marked in $\mc{P}_{i+1}$. Considering the preliminary step of the marking procedure, if a body variable in $\mc{P}_{i}$ appears in the head, the corresponding body variable(s) in $\mc{P}_{i+1}$ also appear in the head of the corresponding rule. Similarly at the propagation step, if a body variable is not in marked positions in head of a rule in $\mc{P}_{i}$, the same variable or its corresponding variables in $\mc{P}_{i+1}$ do not appear in the head of the corresponding rule in $\mc{P}_{i+1}$ in a marked position. Therefore, if a variable is not marked during the steps of the marking procedure in $\mc{P}_{i}$, it is will not get marked in $\mc{P}_{i+1}$ either.
\end{enumerate}

As a result of the claims, if there is no repeated marked variable in $\mc{P}_i$ that  does not appear in finite rank positions, then there is no such variable in $\mc{P}_{i+1}$. Since each iteration preserves the weak-stickiness property, the whole transformation also preserves the property.\boxtheorem}


\vspace{5mm}
\noindent {\bf Proof of Theorem~\ref{th:partial}:} {$\mc{P}'$ is sticky since every weak variable that breaks the weak-stickiness property is replaced with constants. Also, $\nit{ans}(q,\mc{P})=\nit{ans}(q',\mc{P}')$ follows from the correctness of Proposition~\ref{prop:complexity} and the fact that the weak variables are replaced with every possible constant from $D$. It is important that no null value can appear in the positions of the weak variables in $\Sigma_{0,\infty}$ due to the first phase.

The algorithm runs in polynomial time with respect to the size of the database because running the first phase ({\sf ReduceRank}) is data independent and the second phase (partial grounding) replaces variables with polynomially many values from the database.\boxtheorem }


\begin{thebibliography}{10}

\bibitem{abiteboul95}
Abiteboul, S., Hull, R. and Vianu, V.  \newblock {\em Foundations of Databases}.
Addison-Wesley, 1995.

\bibitem{beeri-chase-termination}
Beeri, C. and Vardi, M. Y.
\newblock{The implication problem for data dependencies}.
\newblock{\em Proc. Colloquium on Automata, Languages and Programming}, 1981, pp. 73--85.

\bibitem{beeri}
Beeri, C. and Vardi, M. Y.
\newblock{A Proof Procedure for Data Dependencies}.
\newblock{\em Journal of ACM}, 1984, 31(4):718--741.

\bibitem{cali12-jws}
Cal\`\i, A., Gottlob, G. and Lukasiewicz, T.
\newblock{A General Datalog-based Framework for Tractable Query Answering over Ontologies}.
\newblock{\em Journal of Web Semantics}, 2012, 14:57--83.

\bibitem{cali12}
Cal\`\i, A., Gottlob, G. and Pieris, A.
\newblock{Towards more Expressive Ontology Languages: The Query Answering Problem}.
\newblock{\em Artificial Intelligence}, 2012, 193:87--128.

\bibitem{cali13}
Cal\`\i, A., Gottlob, G. and Kifer, M.
\newblock{Taming the Infinite Chase: Query Answering under Expressive Relational Constraints}.
\newblock{\em Journal of AI Research}, 2013, 48(1):115--174.

\bibitem{deutsch}
Deutsch, A., Nash, A. and Remmel, J. B.
\newblock{The Chase Revisited}.
\newblock{\em Proc. PODS}, 2008, pp. 149–-158.

\bibitem{duschka12}
Duschka, O. M. and Genesereth, M. R.
\newblock{Answering Recursive Queries Using Views}.
\newblock{\em Proc. PODS}, 1997, pp. 109--116.

\bibitem{fagin}
Fagin, R., Kolaitis, P. G., Miller, R. J. and Popa, L.
\newblock{Data Exchange: Semantics and Query Answering}.
\newblock{\em TCS}, 2005, 336:89--124.

\bibitem{gottlob14} Gottlob, G., Orsi, G. and Pieris, A.
\newblock Query Rewriting and Optimization for Ontological Databases.
\newblock {\em Proc. TODS}, 2014, 39(3):25.

\bibitem{johnson}
Johnson, D. S. and Klug, A.
\newblock{Testing Containment of Conjunctive Queries under Functional and Inclusion Dependencies}.
\newblock{\em Proc. PODS}, 1984, pp. 164--169.

\bibitem{krotzsch}
Kr\"{o}tzsch, M. and Rudolph, S.
\newblock{Extending Decidable Existential Rules by Joining Acyclicity and Guardedness}.
\newblock{\em Proc. IJCAI}, 2011, pp. 963--968.

\bibitem{leone11}
Leone, N., Manna, M., Terracina, G. and Veltri, P.
\newblock{Efficiently Computable \datalogpm{} Programs}.  \newblock{\em Proc. KR}, 2012, pp. 13-23.

\bibitem{maier}
Maier, D., Mendelzon, A. and Sagiv, Y.
\newblock{Testing Implications of Data Dependencies}.
\newblock{\em Proc. TODS}, 1979, pp. 152--152.

\bibitem{milani15-amw} Milani, M. and Bertossi, L.
\newblock Tractable Query Answering and Optimization for Extensions of Weakly-Sticky Datalog+-.
\newblock{\em Proc. AMW}, 2015.

\bibitem{MiCB16} Milani, M., Cal\`\i, A. and Bertossi, L.  \newblock A
  Hybrid Approach to Query Answering \\under Expressive Datalog$^\pm$.
  \newblock Tech.~rep.~available at~\texttt{https://goo.gl/edg9FK}.

\end{thebibliography}
\end{document}